\documentclass{PoS}
\usepackage{amsmath}
\renewcommand{\Re}{{\rm Re\thinspace}}
\renewcommand{\Im}{{\rm Im\thinspace}}

\usepackage{diagbox}
\usepackage{ulem}
\def\lsim{\mathrel{\raise.3ex\hbox{$<$\kern-.75em\lower1ex\hbox{$\sim$}}}}

\title{The CP-violating type-II 2HDM and \\
Charged Higgs boson benchmarks}

\ShortTitle{CP-violating type-II 2HDM}

\author{Lorenzo BASSO$^a$, Anna LIPNIACKA$^b$, Farvah MAHMOUDI$^{cd}$,
  Stefano~MORETTI$^{ef}$, \speaker{Per OSLAND}\thanks{The speaker is
    very grateful to George Zoupanos and the Organizing Committee of
    the Corfu Summer Institute 2012 for their invitation and
    hospitality. This research was supported in part by the National
    Science Foundation under Grant No.\ NSF PHY11-25915, by the
    Norwegian Research Council and by the European Community's Seventh
    Framework Programme (FP7/2007-2013) under grant agreement No.\ 290605 (PSI-FELLOW/COFUND).}$^{\ast b}$, Giovanni Marco PRUNA$^g$ and Mahdi~PURMOHAMMADI$^b$ \\ \ \\
\llap{$^a$} Albert-Ludwigs-Universit\"at - Fakult\"at f\"ur Mathematik und Physik, \\ 
        D-79104 Freiburg i.\ Br., Germany \\ 
\llap{$^b$} Department of Physics and Technology, University of Bergen, \\
        Postboks 7803, N-5020 Bergen, Norway\\ 
\llap{$^c$} CERN Theory Division, Physics Department, \\
        CH-1211 Geneva 23, Switzerland \\
\llap{$^d$} Clermont Universit{\'e}, Universit\'e Blaise Pascal, CNRS/IN2P3, LPC, \\ 
        BP 10448, 63000 Clermont-Ferrand, France\\
\llap{$^e$} School of Physics \& Astronomy, University of Southampton, Highfield, \\
        Southampton SO17 1BJ, UK \\ 
\llap{$^f$} Particle Physics Department, Rutherford Appleton Laboratory, \\
        Chilton, Didcot, Oxon OX11 0QX, UK \\
\llap{$^g$}
        Paul Scherrer Institute, CH-5232 Villigen PSI, Switzerland\\  \ \\
        E-mails: \email{Lorenzo.Basso@physik.uni-freiburg.de}, \email{Anna.Lipniacka@ift.uib.no}, \email{Mahmoudi@in2p3.fr}, \email{S.Moretti@soton.ac.uk}, \email{Per.Osland@ift.uib.no}, \email{Giovanni-Marco.Pruna@psi.ch}, \email{Mahdi.PurMohammadi@ift.uib.no}}

\abstract{
We review and update the interpretation of the 125~GeV scalar as the lightest Higgs boson of the Two-Higgs-Doublet Model, allowing for CP violation in the potential. The detection of a charged Higgs boson would exclude the Standard Model. Proposed benchmarks for charged-Higgs searches in the channel $pp\to H^+W^-X\to W^+W^-H_1X$ are reviewed and updated.
}

\FullConference{Proceedings of the Corfu Summer Institute 2012 "School and Workshops on Elementary Particle Physics and Gravity"\\
		September 8-27, 2012\\
		Corfu, Greece}

\begin{document}
\section{Introduction and notation}
One of the most pressing questions in particle physics today is to determine whether or not the discovered Higgs-like particle \cite{:2012gk,:2012gu}
is compatible with the Standard-Model expectations \cite{Djouadi:2005gi}, or whether it belongs to an enlarged scalar sector.
The most popular extension is certainly the supersymmetric one \cite{Djouadi:2005gj}, but while waiting for hints of supersymmetry, it may be worthwhile to entertain also other possibilities \cite{Gunion:1989we}.

We shall here review and update the interpretation of the 125~GeV Higgs particle in the Two-Higgs-Doublet model \cite{Basso:2012st}, with Type~II Yukawa couplings, like in the MSSM.
In the spirit of the original motivation for the model, we allow CP violation \cite{Lee:1973iz} (see also Ref.~\cite{Shu:2013uua}), and take the potential to be
\begin{align}
\label{Eq:pot_7}
V&=\frac{\lambda_1}{2}(\Phi_1^\dagger\Phi_1)^2
+\frac{\lambda_2}{2}(\Phi_2^\dagger\Phi_2)^2
+\lambda_3(\Phi_1^\dagger\Phi_1) (\Phi_2^\dagger\Phi_2) \nonumber \\
&+\lambda_4(\Phi_1^\dagger\Phi_2) (\Phi_2^\dagger\Phi_1)
+\frac{1}{2}\left[\lambda_5(\Phi_1^\dagger\Phi_2)^2+{\rm h.c.}\right] \\
&-\frac{1}{2}\left\{m_{11}^2(\Phi_1^\dagger\Phi_1)
\!+\!\left[m_{12}^2 (\Phi_1^\dagger\Phi_2)\!+\!{\rm h.c.}\right]
\!+\!m_{22}^2(\Phi_2^\dagger\Phi_2)\right\}. \nonumber
\end{align}
with $m_{12}^2$ and $\lambda_5$ complex.

The potential uniquely determines the mass spectrum. There are three neutral states, $H_1$, $H_2$ and $H_3$, with masses $M_1\leq M_2\leq M_3$, and a charged pair, $H^\pm$, with mass $M^\pm$.
With the field decomposition (ghosts are removed):
\begin{equation}\label{Eq:Higgs_goldstones}
\Phi_1=
\left(
\begin{array}{c}
- s_\beta H^+ \\
\frac{1}{\sqrt{2}} [v_1 + \eta_1 - i s_\beta \eta_3]
\end{array}
\right), \qquad
\Phi_2 =
\left(
\begin{array}{c}
c_\beta H^+ \\
\frac{1}{\sqrt{2}} [v_2 + \eta_2 + i c_\beta \eta_3 ]
\end{array}
\right).
\end{equation}
where $c_\beta=\cos\beta$ and $s_\beta=\sin\beta$, and the ratio defines
$\tan{\beta}=v_2/v_1$,
the neutral Higgs states can be expressed via a rotation matrix $R$ as
\begin{equation} \label{Eq:R-def}
\begin{pmatrix}
H_1 \\ H_2 \\ H_3
\end{pmatrix}
=R
\begin{pmatrix}
\eta_1 \\ \eta_2 \\ \eta_3
\end{pmatrix}.
\end{equation}
The rotation matrix can be parametrized in terms of 3 rotation angles, $\alpha_j$ \cite{Accomando:2006ga}.

It is instructive to take the input parameters in terms of quantities that have a more direct physical interpretation.
A convenient choice is to specify the masses of the two lightest Higgs bosons ($M_1$, $M_2$), as well as that of the charged one, $M^\pm$. Supplementing these data by $\tan\beta$, the three angles $\alpha_j$, as well as $\mu^2=\Re m_{12}^2/(2\cos\beta\sin\beta)$, the potential can be trivially reconstructed \cite{Khater:2003wq,ElKaffas:2007rq}. This is useful, since some of the theoretical constraints (see below) are more simply expressed in terms of the potential parameters.

\section{Constraints}
The parameter space is constrained both by theoretical considerations, and by experimental data.
\subsection{Theory constraints}
We impose the familiar theory constraints: positivity, tree-level unitarity, and perturbativity. In addition, we impose the less familiar constraint of requiring that the potential minimum be a global one. This is computationally rather expensive, and therefore checked only if all other constraints (including experimental ones) are satisfied. For details, see Ref.~\cite{Basso:2012st}.
\subsection{Experimental constraints}
We impose the constraints from flavor physics (in particular, $b\to s\gamma$), $\Gamma(Z\to b\bar b)$, and electroweak precision observables $T$ and $S$.
While allowing for CP violation opens up a larger parameter space than CP-conserving models, one has to make sure that excessive CP violation is not induced. A representative observable that is easily checked, is the electron electric dipole moment. For details, see Ref.~\cite{Basso:2012st}.

As compared with our original work, the LHC constraints have tightened: ATLAS has presented the new (preliminary) result $R_{\gamma\gamma}=1.65\pm0.3$ \cite{ATLAS:2013oma}, where
\begin{equation} \label{Eq:R_gammagamma}
R_{\gamma\gamma}=\frac{\sigma(pp\to H_1X){\rm BR}(H_1\to\gamma\gamma)}
{\sigma(pp\to H_\text{SM}X){\rm BR}(H_\text{SM}\to\gamma\gamma)}.
\end{equation}
As a 2-$\sigma$ interval, we allow $1.05\leq R_{\gamma\gamma}\leq2.33$. The exclusion of values below unity has significant implications for the allowed parameter range.\footnote{After this scan was performed, also CMS released their updated results for $R_{\gamma\gamma}$ \cite{CMS-2013}. They find the 2-$\sigma$ range $0.26\leq R_{\gamma\gamma}\leq1.34$, significantly lower than that obtained by ATLAS, and adopted here. Since the ATLAS and CMS results barely overlap, our scans should not be taken as definitive, but rather as an illustration of how improved data can further constrain the model.}

Also, CMS has presented more tight exclusions of a SM-like Higgs particle in the high-mass region \cite{Guillelmo:2013cca},
\begin{equation} \label{Eq:R_ZZ}
R_{ZZ}=\frac{\sigma(pp\to H_jX)){\rm BR}(H_j\to ZZ)}
{\sigma(pp\to H_\text{SM}X){\rm BR}(H_\text{SM}\to ZZ)},
\end{equation}
having implications for how strongly the heavier partners, $H_2$ and $H_3$ can couple to $W$ and $Z$. Also, both ATLAS \cite{ATLAS:2013oma} and CMS \cite{Guillelmo:2013cca} presented new results on $R_{ZZ}$ and $R_{WW}$, relevant for $H_1$. As a 2-$\sigma$ envelope covering both the $ZZ$ and $WW$ channels, we adopt the range $0.3\leq R_{VV}\leq2.7$. Finally, we also take into account new preliminary results from a $H\to b\bar b$ search \cite{ATLAS:2013nma}, yielding the 2-$\sigma$ range $0.49\leq R_{b\bar b}\leq 1.69$, with $R_{b\bar b}$ defined in analogy with (\ref{Eq:R_gammagamma}) and (\ref{Eq:R_ZZ}).

In both (\ref{Eq:R_gammagamma}) and (\ref{Eq:R_ZZ}), we approximate the cross section ratio by the corresponding ratio of gluon-gluon branching ratios,
\begin{equation}
\frac{\sigma(pp\to H_jX)}{\sigma(pp\to H_\text{SM}X)}
\simeq\frac{\Gamma(H_j\to gg)}{\Gamma(H_\text{SM}\to gg)},
\end{equation}
i.e., we consider only production via the dominant gluon-gluon fusion.

It is interesting to see how the new data have led to a shrinking of the allowed parameter space. We shall refer to the constraints described above as ``Moriond 2013'' and compare the still allowed parameter space to that allowed by the constraints considered in Ref.~\cite{Basso:2012st}, which we refer to as ``2012''. Since the constraints on $R_{VV}$ and $R_{b\bar b}$ are still rather loose, the main impact of the new data are in the following two sectors: (1) tighter constraints on how $H_2$ and $H_3$ couple to $ZZ$ (or $W^+W^-$), and (2) tighter constraints on how $H_1$ couples to photons, $R_{\gamma\gamma}$.

Key features of these LHC constraints can be summarized as follows:
\begin{alignat}{4} \label{Eq:constraint-2012}
&\text{``2012'' \cite{Basso:2012st}:} &\qquad
H_1\to\gamma\gamma: 0.5\leq &R_{\gamma\gamma}\leq2, &\qquad
&H_{2,3}\to VV:&\quad
&\text{Ref.}~\cite{ATLAS:2012ae,Chatrchyan:2012tx}\\
&\text{``Moriond 2013'':} &\qquad
H_1\to\gamma\gamma: 1.03\leq &R_{\gamma\gamma}\leq2.33, &\qquad
&H_{2,3}\to VV:&\quad
&\text{Ref.}~\cite{ATLAS:2013oma,Guillelmo:2013cca}
\label{Eq:constraints-2013}
\end{alignat}

\section{Allowed parameter space}
We shall here illustrate how the allowed parameter space has shrunk as a result of the recent LHC results, and determine remaining regions.

\subsection{The general CP-violating case}
The $H_1t\bar t$ coupling is essential to the production of $H_1$ via gluon-gluon fusion.
In the general case, the $H_it\bar t$ coupling differs from that of the SM by the factor
\begin{equation}  \label{Eq:H_j_Yuk}
H_j  t\bar t \sim
\frac{1}{\sin\beta}\, [R_{j2}-i\gamma_5\cos\beta R_{j3}],
\end{equation}
where $R_{jk}$ refers to the matrix defined by Eq.~(\ref{Eq:R-def}).
Two points are worth noting: (1) At low $\tan\beta$, a reduced value of $|R_{12}|=|\sin\alpha_1\cos\alpha_2|$ can be compensated for by the factor $\sin\beta$ in the denominator, and (2) there is an additional contribution from the pseudoscalar coupling, proportional to $R_{13}/\tan\beta$ (and a different loop function).

For the $H_1\to\gamma\gamma$ rate, the $H_1W^+W^-$ coupling is likewise essential. The $H_jZZ$ (and $H_jW^+W^-$) coupling is, relative to that of the SM, given by
\begin{equation} \label{Eq:ZZH}
H_j ZZ\  (H_jW^+W^-)\sim
[\cos\beta R_{j1}+\sin\beta R_{j2}].
\end{equation}
This coupling, for $H_2$ and $H_3$, is also important for the constraint from the high-mass exclusion of a Higgs particle, since the most sensitive channels are various sub-channels of the $H\to ZZ$ and $H\to W^+W^-$ ones.

\begin{figure}[htb] 
\begin{center}
 \includegraphics[width=0.85\textwidth]{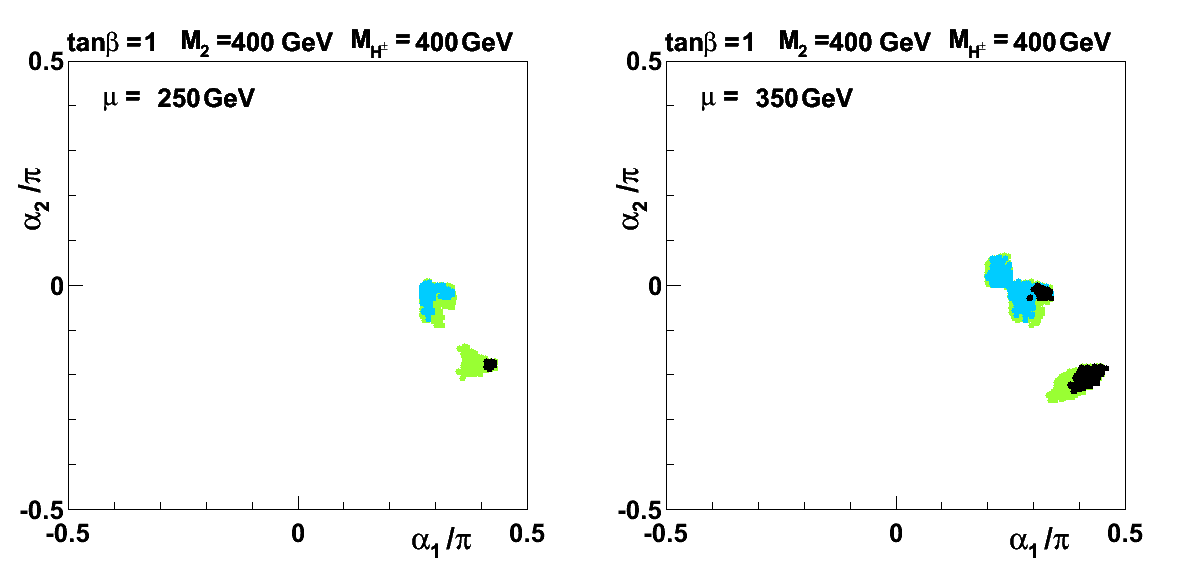}
 \caption{Allowed regions in the $\alpha_1$--$\alpha_2$
  parameter space, with ``2012'' (green) and ``Moriond 2013'' constraints (blue and black), for
  $\tan\beta=1$, $M_2=400~\text{GeV}$ and $M_{H^\pm}=400~\text{GeV}$.}
\label{Fig:alphas12-01-400-400}
\end{center}
\end{figure}

\begin{figure}[htb] 
\begin{center}
 \includegraphics[width=0.85\textwidth]{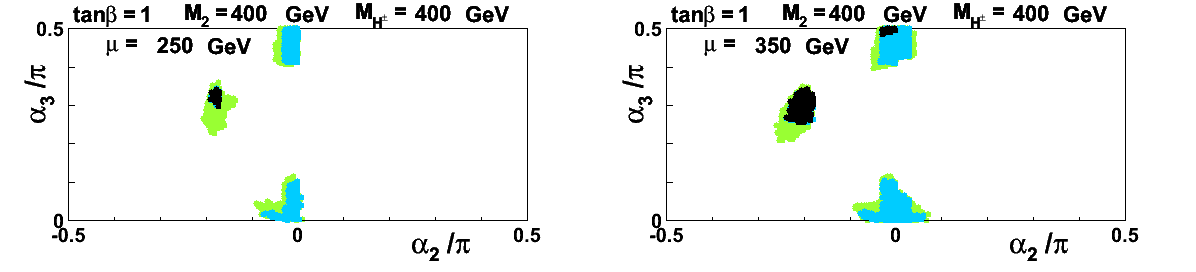}
 \caption{Allowed regions in the $\alpha_2$--$\alpha_3$
  parameter space, with ``2012'' (green) and ``Moriond 2013'' constraints (blue and black), for
  $\tan\beta=1$, $M_2=400~\text{GeV}$ and $M_{H^\pm}=400~\text{GeV}$.}
\label{Fig:alphas23-01-400-400}
\end{center}
\end{figure}

We shall show a few scans over the $\alpha$-space, identifying allowed regions. We start with the case
\begin{equation} \label{Eq:parameters-01-400-400}
\tan\beta=1, \quad M_2=400~\text{GeV}, \quad M_{H^\pm}=400~\text{GeV},
\end{equation}
and two values of $\mu$, namely $\mu=250~\text{GeV}$ and $\mu=350~\text{GeV}$.
In Fig.~\ref{Fig:alphas12-01-400-400} we show allowed regions in the $\alpha_1$--$\alpha_2$ plane, whereas in Fig.~\ref{Fig:alphas23-01-400-400} we show allowed regions in the $\alpha_2$--$\alpha_3$ plane.
We see that the new constraints permit solution for both values of $\mu$ (for these values of $\tan\beta$, $M_2$ and $M_{H^\pm}$). However, the case of $\mu=200~\text{GeV}$, studied in Ref.~\cite{Basso:2012st}, is no longer allowed. This is due to the new constraint on $R_{\gamma\gamma}$.

Already within the ``2012'' constraints, the parameter space is very constrained, as shown in green (in some cases, this is practically covered by the blue and black regions). When we impose the constraints from the high-mass exclusion (reduced couplings of $H_2$ and $H_3$), we obtain the blue regions. 
When we also impose the $R_{\gamma\gamma}$ constraint, we are left with the black regions.

In these examples, there are three regions that are almost or fully allowed (see Fig.~\ref{Fig:alphas23-01-400-400}). There are two regions at small values of $\alpha_2$, one around $\alpha_3=0$, and the other around $\alpha_3=\pi/2$. These both have $\alpha_1\sim\pi/4$. The third region has $\alpha_2<0$, somewhat larger values of $\alpha_1$, and $\alpha_3\sim\pi/4$.

The two first-mentioned regions are close to CP-conserving limits. We recall that $H_3=A$ corresponds to $(\alpha_2,\alpha_3)=(0,0)$, whereas $H_2=A$ corresponds to $(\alpha_2,\alpha_3)=(0,\pi/2)$ \cite{ElKaffas:2007rq}, $A$ being the CP-odd boson. 

\begin{figure}[htb] 
\begin{center}
 \includegraphics[width=0.85\textwidth]{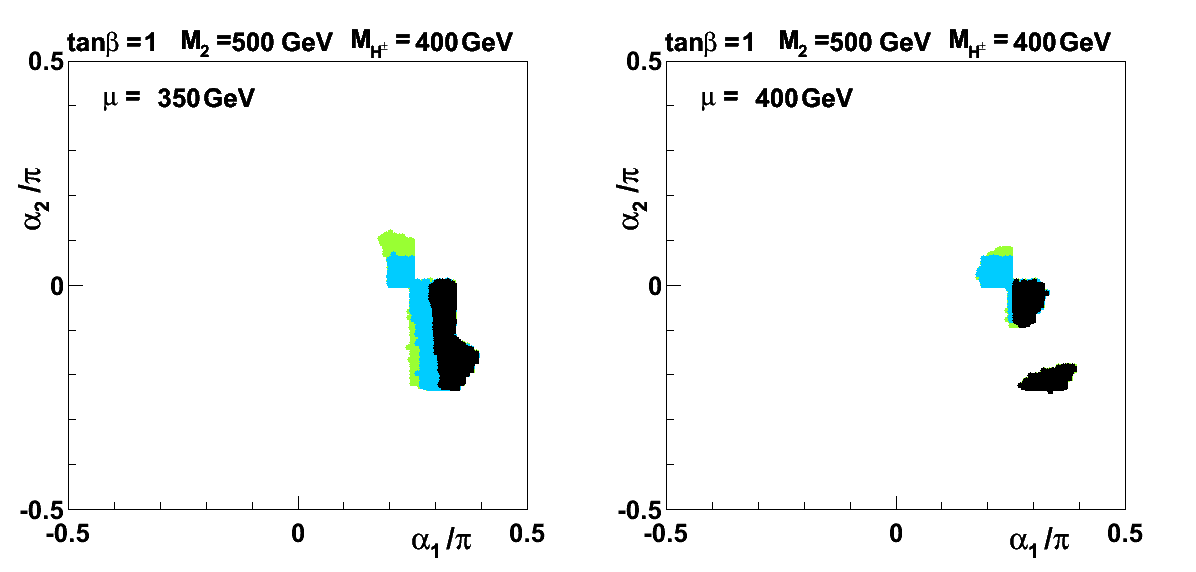}
 \caption{Allowed regions in the $\alpha_1$--$\alpha_2$
  parameter space, with ``2012'' (green) and ``Moriond 2013'' constraints (blue and black), for
  $\tan\beta=1$, $M_2=500~\text{GeV}$ and $M_{H^\pm}=400~\text{GeV}$.}
\label{Fig:alphas12-01-500-400}
\end{center}
\end{figure}

\begin{figure}[htb] 
\begin{center}
 \includegraphics[width=0.85\textwidth]{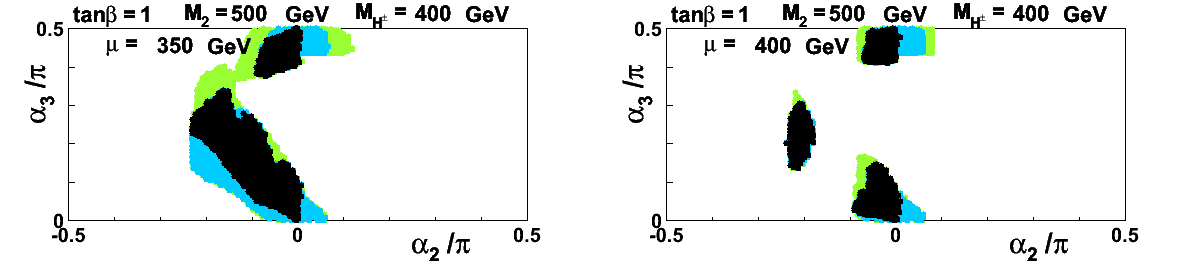}
 \caption{Allowed regions in the $\alpha_2$--$\alpha_3$
  parameter space, with ``2012'' (green) and ``Moriond 2013'' constraints (blue and black), for
  $\tan\beta=1$, $M_2=450~\text{GeV}$ and $M_{H^\pm}=400~\text{GeV}$.}
\label{Fig:alphas23-01-500-400}
\end{center}
\end{figure}

As a second example, we consider 
\begin{equation} \label{Eq:parameters-01-500-400}
\tan\beta=1, \quad M_2=500~\text{GeV}, \quad M_{H^\pm}=400~\text{GeV},
\end{equation}
and show allowed regions in Figs.~\ref{Fig:alphas12-01-500-400} and \ref{Fig:alphas23-01-500-400} for two values of $\mu$, this time 350 and 400~GeV.

For the lower value of $\mu$, two of the three regions described above, are connected. In fact, a large region of CP-violating parameter sets is allowed. For the higher value of $\mu$, all three regions are allowed, but distinct.

\begin{figure}[htb] 
\begin{center}
 \includegraphics[width=0.85\textwidth]{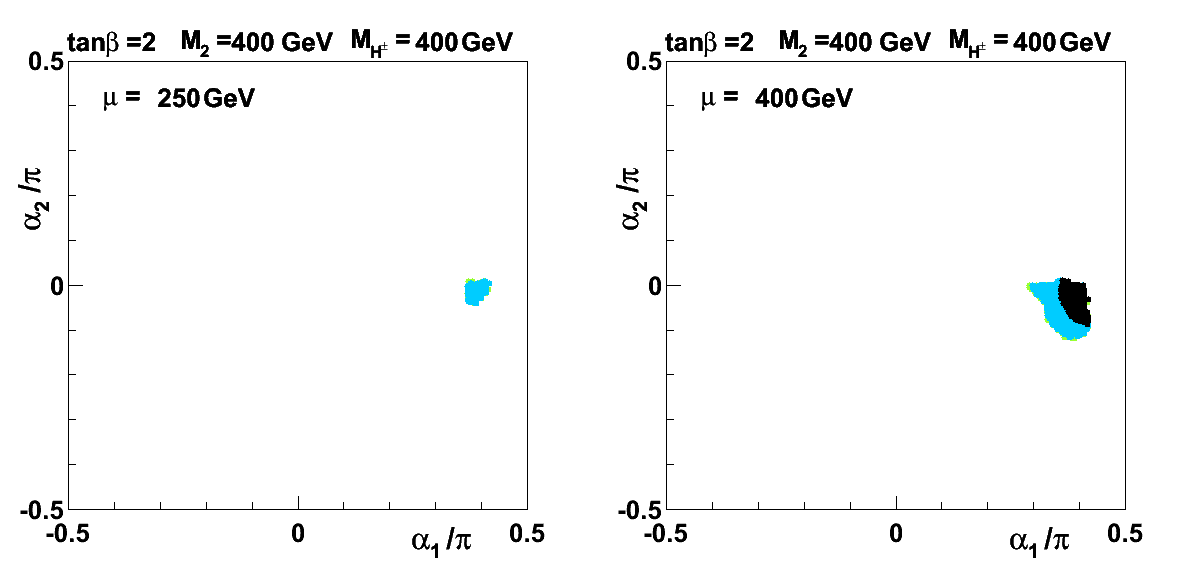}
 \caption{Allowed regions in the $\alpha_1$--$\alpha_2$
  parameter space, with ``2012'' (green) and ``Moriond 2013'' constraints (blue and black), for
  $\tan\beta=2$, $M_2=400~\text{GeV}$ and $M_{H^\pm}=400~\text{GeV}$.}
\label{Fig:alphas12-02-400-400}
\end{center}
\end{figure}

\begin{figure}[htb] 
\begin{center}
 \includegraphics[width=0.85\textwidth]{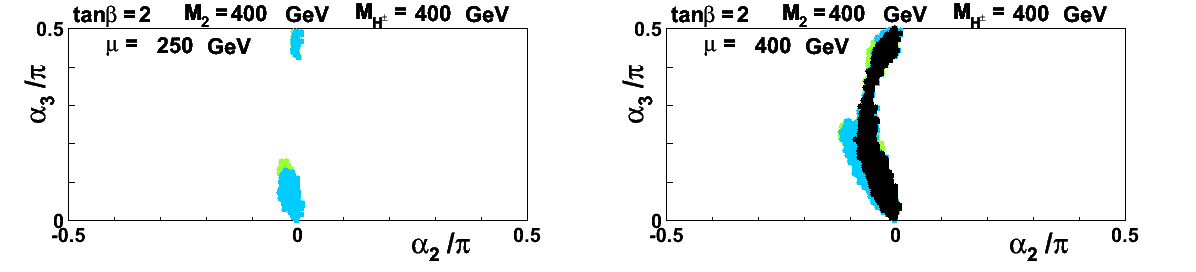}
 \caption{Allowed regions in the $\alpha_2$--$\alpha_3$
  parameter space, with ``2012'' (green) and ``Moriond 2013'' constraints (blue and black), for
  $\tan\beta=2$, $M_2=400~\text{GeV}$ and $M_{H^\pm}=400~\text{GeV}$.}
\label{Fig:alphas23-02-400-400}
\end{center}
\end{figure}

As a third example, we consider $\tan\beta=2$,  with $M_2=M_{H^\pm}=400~\text{GeV}$, in Figs.~\ref{Fig:alphas12-02-400-400} and \ref{Fig:alphas23-02-400-400}. Here, we take $\mu=250~\text{GeV}$ and $\mu=400~\text{GeV}$.
In this case, the blue regions practically cover the green regions, meaning that the ``Moriond 2013'' constraints in the high-mass region have little impact. Instead, the new $R_{\gamma\gamma}$ constraint significantly reduces the allowed regions (black). Indeed, the low-$\mu$ case is fully excluded. For $\mu=400~\text{GeV}$, there is a major difference with respect to the $\tan\beta=1$ case: the whole range of $\alpha_3$-values is allowed, i.e., in addition to the CP-conserving limits also a band in the CP-violating interior of the $\alpha_2$--$\alpha_3$ space is allowed.

In Tables~\ref{Table:tanbeta01-m2-mch-mu} and \ref{Table:tanbeta02-m2-mch-mu} we summarize the results of very coarse scans over $\mu$, for selected grids in $M_2$ and $M_{H^\pm}$. Centered roughly around the average of these values, 50~GeV increments in $\mu$ are explored and reported in these tables. For example, the notation ``[250,350]'' means that $\mu$ values of 250, 300 and 350 give allowed solutions, whereas the next values below (200) and above (400) do not.

\newcommand{\twolines}[2]{\genfrac{}{}{0pt}{}{\text{#1}}{\text{#2}}}
\begin{table}[ht]
\begin{center}
\begin{tabular}{|c|c|c|c|c|c|}
\hline
\diagbox{$M_{H^\pm}$}{$M_2$} & 300 & 350 & 400 & 450 & 500  \\
\hline
 500 & none &  none & none & none & [450,500] \\
\hline
 450 & none &  [250,350] & [300,400] & [350,450] & [400,450] \\
\hline
 400 & none &  [250,350] & [250,400] & [300,400] & [350,400] \\
\hline
\end{tabular}
\end{center}
\caption{Some allowed values of $\mu$ for selected values of $M_2$ and $M_{H^\pm}$ [all in GeV], for $\tan\beta=1$.  \label{Table:tanbeta01-m2-mch-mu}}
\end{table}

\begin{table}[ht]
\begin{center}
\begin{tabular}{|c|c|c|c|c|c|}
\hline
\diagbox{$M_{H^\pm}$}{$M_2$} & 300 & 350 & 400 & 450 & 500  \\
\hline
 500 & none &  none & none & none & [500] \\
\hline
 450 & none &  none & [400] & [400,450] & [450,500] \\
\hline
 400 & none &  [350] & [350,450] & [350,450]  & [400] \\
\hline
\end{tabular}
\end{center}
\caption{Some allowed values of $\mu$ for selected values of $M_2$ and $M_{H^\pm}$ [all in GeV], for $\tan\beta=2$.  \label{Table:tanbeta02-m2-mch-mu}}
\end{table}

We see that the allowed range of $\mu$ is rather narrow, and constrained to lie around ${\cal O}(M_2,M_{H^\pm})$. See also the discussion in sect.~\ref{Sect:high-M-tanbeta}.

Related aspects of the model were presented recently \cite{Barroso:2013zxa}. We agree with the findings of those authors that it is difficult for this model to yield high values of $R_{\gamma\gamma}$ consistent with the other constraints.


\subsection{The CP-conserving case}
\begin{figure}[htb] 
\begin{center}
 \includegraphics[width=0.85\textwidth]{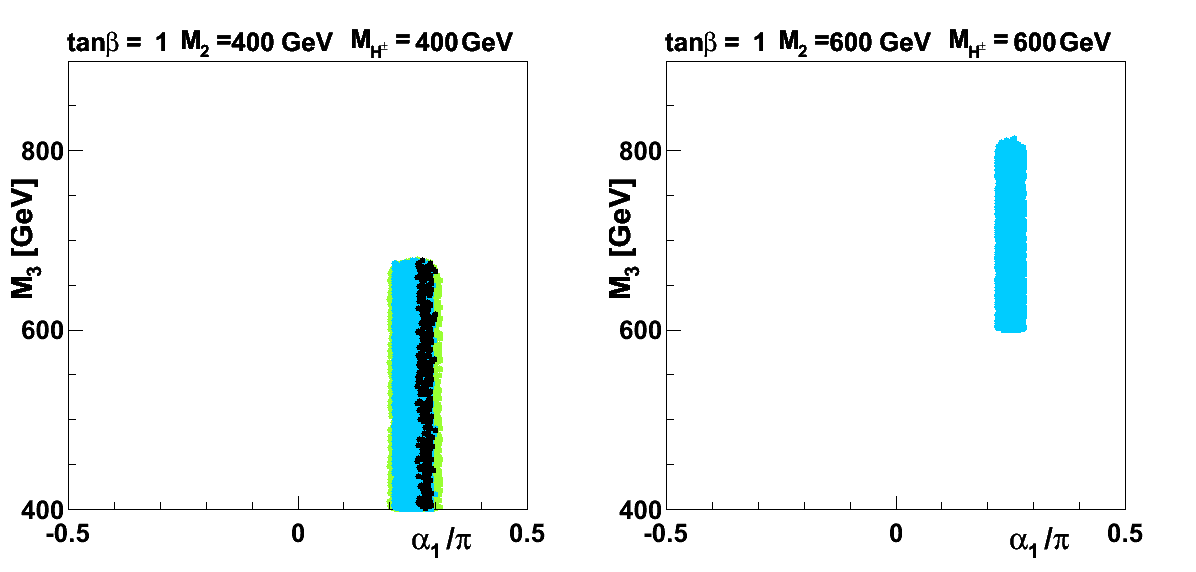}
 \caption{Allowed regions in the $\alpha_1$--$M_3$
  parameter space, with ``2012'' (green) and ``Moriond 2013'' constraints (blue and black), for
  $\tan\beta=1$, $M_2=400~\text{GeV}$ and 600~GeV.}
\label{Fig:alphas1-mh3-01}
\end{center}
\end{figure}

\begin{figure}[htb] 
\begin{center}
 \includegraphics[width=0.85\textwidth]{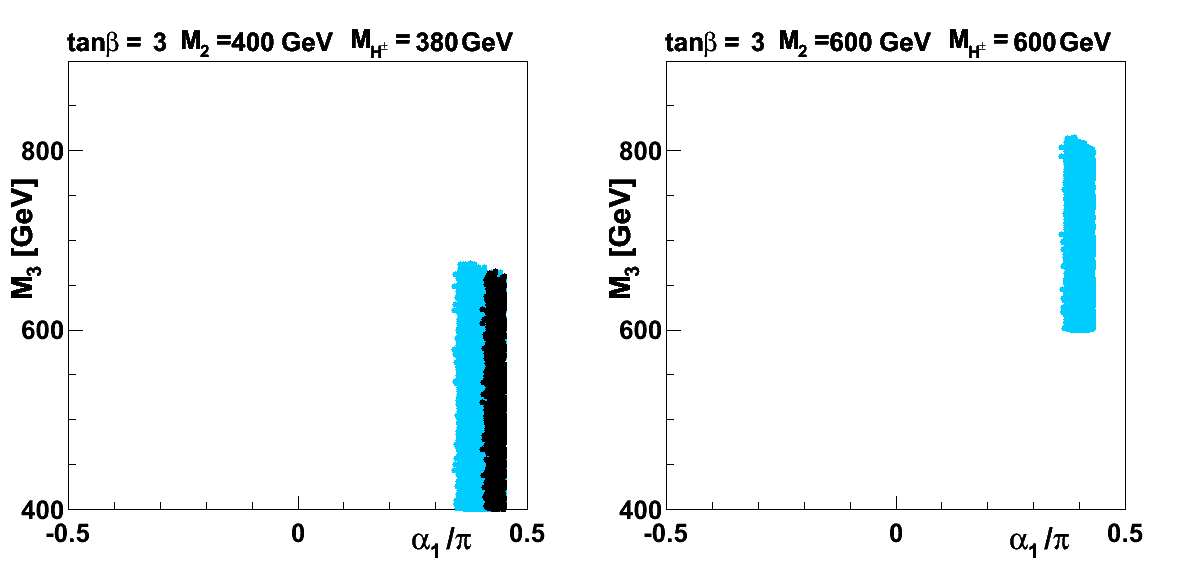}
 \caption{Allowed regions in the $\alpha_1$--$M_3$
  parameter space, with ``2012'' (green) and ``Moriond 2013'' constraints (blue and black), for
  $\tan\beta=3$, $M_2=400~\text{GeV}$ and 600~GeV.}
\label{Fig:alphas1-mh3-03}
\end{center}
\end{figure}

There are three CP-conserving limits, any one of $H_1$, $H_2$ or $H_3$ (with $M_1\leq M_2\leq M_3$) could be CP-odd, usually referred to as $A$. We shall here discuss the case $H_3=A$, which in the above terminology corresponds to $\alpha_2=0$, $\alpha_3=0$, namely the origin in Figs.~\ref{Fig:alphas23-01-400-400}, \ref{Fig:alphas23-01-500-400} and \ref{Fig:alphas23-02-400-400}.
In Figs.~\ref{Fig:alphas1-mh3-01} and \ref{Fig:alphas1-mh3-03} we show allowed regions in the $\alpha_1$-$M_3$ plane, in the same color code as above. Two values of $\tan\beta$ are studied (1 and 3), and two values of $M_2$ (400 and 600~GeV).

In this limit, the $H_iZZ$ (or $H_iW^+W^-$) couplings are proportional to
\begin{equation}
H_1ZZ\sim\cos(\beta-\alpha_1), \quad
H_2ZZ\sim\sin(\beta-\alpha_1), \quad
H_3ZZ=0,
\end{equation}
(in the familiar notation of the CP-conserving model, $\cos(\beta-\alpha_1)\to\sin(\beta-\alpha)$ and $\sin(\beta-\alpha_1)\to-\cos(\beta-\alpha)$).
Thus, maximizing the $H_1W^+W^-$ coupling (in order to obtain an acceptable $H_1\to\gamma\gamma$ rate), simultaneously makes the $H_2ZZ$ coupling small, and the tightened ``Moriond 2013'' high-mass exclusion has little effect.
In fact, for the CP-conserving case, it is only at low $\tan\beta$ and low $M_2$ (see left panel of Fig.~\ref{Fig:alphas1-mh3-01}) that they have any impact. For higher values of $\tan\beta$ and $M_2$, even rather loose constraints (\ref{Eq:constraint-2012}) on $H_1\to\gamma\gamma$ are more relevant than the high-mass ``Moriond 2013'' exclusion applied to $H_2$ and $H_3$.

For low values of $M_2$, the new $R_{\gamma\gamma}$ constraints significantly reduce the allowed range in $\alpha_1$, whereas for higher values of $M_2$ everything is excluded.

The CP-conserving case was recently studied in \cite{Grinstein:2013npa,Coleppa:2013dya,Chen:2013rba,Eberhardt:2013uba}. Our results are in qualitative agreement with one notable exception: Eberhardt {\it et al.} \cite{Eberhardt:2013uba} find that high masses (of order 1~TeV)  are allowed, whereas we do not. We comment on that limit in Sect.~\ref{Sect:high-M-tanbeta}.
\begin{figure}[htb]
\begin{center}
 \includegraphics[width=0.7\textwidth]{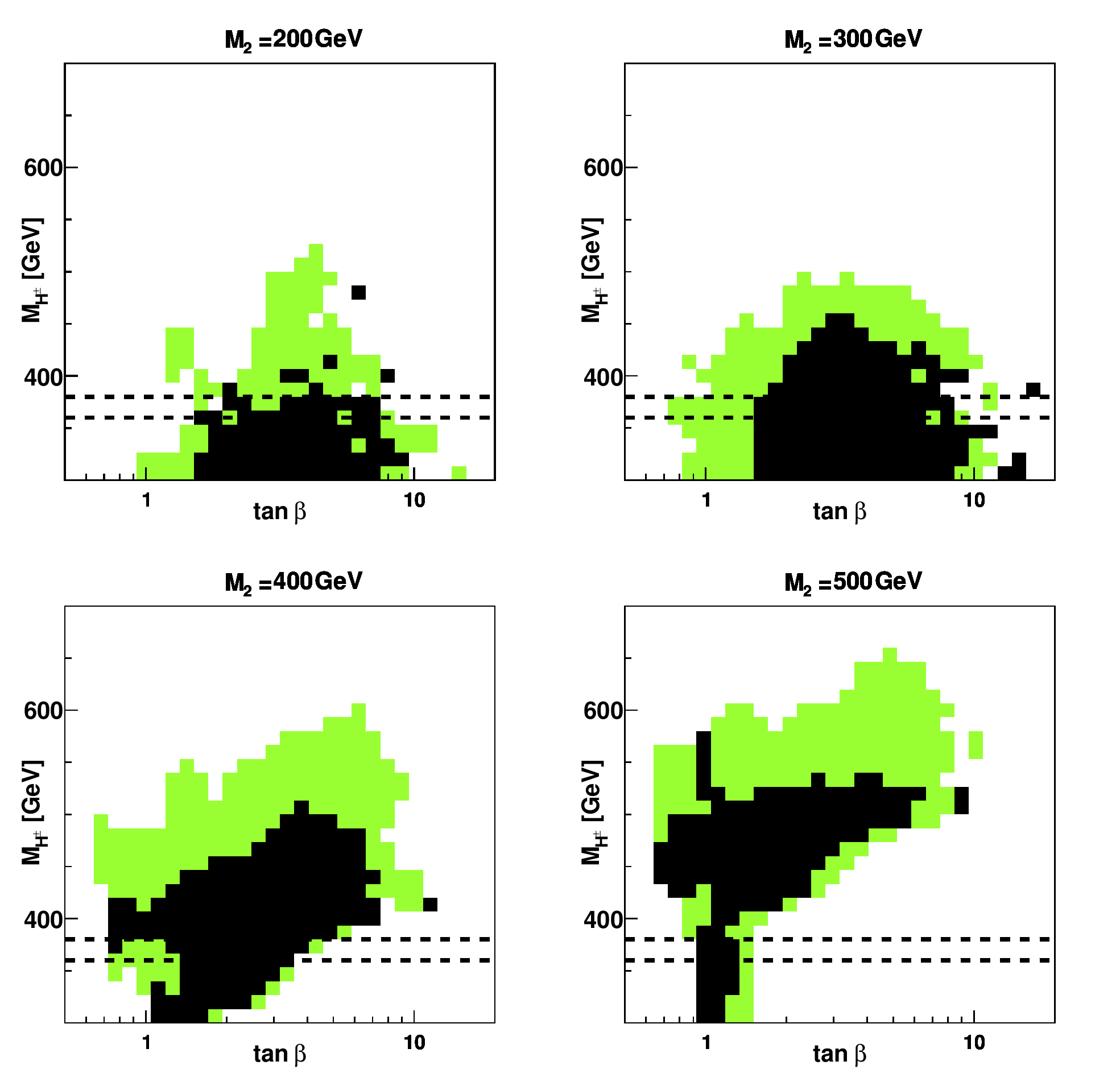}
\caption{Allowed regions in the $\tan\beta$--$M_{H^\pm}$
  parameter space, without (green) and with (black) the recent LHC constraints, for $M_1=125~\text{GeV}$
  and four values of $M_2$,
  as indicated.
The dashed lines show the recent lower bounds at 360 and 380~GeV \cite{Hermann:2012fc}.
\label{Fig:tanbeta-mh_ch}}
\end{center}
\end{figure}

\subsection{Overview vs $\tan\beta$ and $M_{H^\pm}$}

In Fig.~\ref{Fig:tanbeta-mh_ch} we present an overview, in the $\tan\beta$-$M_{H^\pm}$ plane, of allowed regions. Two colors are used: green refers to regions considered allowed in our 2012 paper \cite{Basso:2012st}, whereas black refers to regions compatible with the recent ``Moriond 2013'' constraints (adopting the ATLAS range for $R_{\gamma\gamma}$). The horizontal dashed lines represent the recent constraint from $b\to s\gamma$ transitions \cite{Hermann:2012fc}, according to which, $M_{H^\pm}<360$ or $380~\text{GeV}$ is excluded for all $\tan\beta$. (The published version of the paper quotes both these values, depending on the choice of input.)

The new constraints from LHC are seen to significantly reduce the allowed region in parameter space, in particular at high values of $M_{H^\pm}$. However, we note that they depend on adopting the ATLAS result for $R_{\gamma\gamma}$ \cite{ATLAS:2013oma}.

\section{Charged Higgs Benchmarks}
We consider the production of the charged Higgs boson in association with a $W$ boson, and its decay to a neutral Higgs boson and a second $W$:
\begin{equation}
pp\to W^\mp H^\pm \to W^\mp W^\pm H_1 \to W^\mp W^\pm b \bar{b} \to 2j+2b+1\ell + \mbox{MET}.
\end{equation}
i.e., we let one $W$ decay hadronically, and the other leptonically. 

There is a considerable $t\bar t$ background, but it was found \cite{Basso:2012st} that a certain combination of cuts can reduce this background to a tolerable level (see also Ref.~\cite{Basso:2013hs}).

In Ref.~\cite{Basso:2012st} we proposed a set of benchmarks, which, in
addition to being compatible with the constraints, also yielded
acceptable signal rates for the above channel. In the face of the new
LHC constraints, they are {\it all excluded}. Then, in the spirit of our previous work, we have adopted the same procedure and picked new benchmarks which could replace the old ones with no qualitative modification of our previous strategy to clean the signal from the background. In Table~\ref{Table:points} we present a new set of candidate points: the value of $\tan{\beta}$ is chosen to be $\mathcal{O}(1)$ and the free mass parameters are $\sim 400-500$ GeV.

\begin{table}[ht]
\begin{center}
\begin{tabular}{|c|c|c|c|c|c|c|}
\hline
\ & $\alpha_1/\pi$ & $\alpha_2/\pi$ & $\alpha_3/\pi$ & $\tan\beta$ & $M_2$ & $M_{H^\pm}^\text{min} ,M_{H^\pm}^\text{max}$ \\
\hline
 $P_1^\prime$ & $0.27$ & $-0.002$ & $0.045$ & $1$ & $400$ & $\sim400$ \\
 $P_2^\prime$ & $0.27$ & $-0.02$ & $0.02$  & $1$ & $400$ & $\sim400$ \\
 $P_3^\prime$ & $0.29$ & $-0.03$ & $0.03$  & $1$ & $450$ & $\sim400$ \\
 $P_4^\prime$ & $0.38$ & $-0.22$ & $0.26$  & $1$ & $450$ & $\sim450$ \\
 $P_5^\prime$ & $0.37$ & $-0.15$ & $0.23$  & $1$ & $500$ & $\sim400$ \\
 $P_6^\prime$ & $0.38$ & $-0.02$ & $0.04$  & $2$ & $400$ &$\sim400$ \\
 $P_7^\prime$ & $0.39$ & $-0.03$ & $0.13$  & $2$ & $400$ & $\sim400$ \\
 $P_8^\prime$ & $0.38$ & $-0.0005$ & $0.49$  & $2$ & $400$ & $\sim400$ \\
\hline
\end{tabular}
\end{center}
\caption{New benchmark points selected from the allowed parameter space. Masses $M_2$ and allowed range of $M_{H^\pm}$ are in GeV, and $\mu\simeq\min(M_2,M_{H^\pm})$.  \label{Table:points}}
\end{table}

All the candidate points are allowed by the theoretical and experimental contraints that we have described in the previous sections, but the open question is whether at these benchmarks our selection strategy is still allowed. In order to establish this, we must profile the charged Higgs at each point of the parameter space and check if there is room for an observation at the LHC. In Table~\ref{Table:BRHc} we show the branching ratios for the charged Higgs boson main decay channels: $WH_1$, $tb$ and $ts$. Since we are interested in decay and production associated with bosons, we focus on the $WH_1$ decay mode. From the table is clear that the points $P'_4$, $P'_5$ and $P'_7$ provide interesting branching ratio values. 

\begin{table}[ht]
\begin{center}
\begin{tabular}{|c|c|c|c|c|c|c|c|c|}
\hline
\ & $P_1^\prime$ & $P_2^\prime$ & $P_3^\prime$ & $P_4^\prime$ & $P_5^\prime$ & $P_6^\prime$ & $P_7^\prime$ & $P_8^\prime$ \\
\hline
 $W^+ H_1$ & $0.0034$ & $0.0067$ & $0.020$ & $0.35$ &
 $0.21$ & $0.037$ & $0.071$ & $0.025$ \\
 $t\bar{b}$ & $0.99$ & $0.99$ & $0.98$  & $0.64$ &
 $0.79$ & $0.96$ & $0.93$ & $0.97$ \\
  $t\bar{s}$ & $0.0016$ & $0.0016$ & $0.0016$  & $0.0010$ &
 $0.0012$ & $0.0015$ & $0.0015$ & $0.0015$ \\
\hline
\end{tabular}
\end{center}
\caption{Branching ratios of the main charged Higgs decay channels. \label{Table:BRHc}}
\end{table}

Thereafter, in Table~\ref{Table:CSHc} we present the cross sections for the main charged Higgs production channels at the LHC with $\sqrt{s}=8$ TeV and $\sqrt{s}=14$ TeV. Again, we are interested in the production associated with bosons, in particular the most sizeable one, which is $H^\pm W^\mp$. By direct comparison, we immediately see that the most promising points are represented by the choice $P'_1$, $P'_2$,  $P'_3$, $P'_5$ and $P'_6$.

\begin{table}[ht]
\begin{center}
\begin{tabular}{|c|c|c|c|c|c|c|c|c|}
\hline
\ & $P_1^\prime$ & $P_2^\prime$ & $P_3^\prime$ & $P_4^\prime$ & $P_5^\prime$ & $P_6^\prime$ & $P_7^\prime$ & $P_8^\prime$ \\
\hline
 $H^\pm H_i (8)$ & $0.0010$ & $0.00099$ & $0.00073$ & $0.0013$ &
 $0.0013$ & $0.00075$ & $0.00092$ & $0.0010$ \\
 $H^\pm H_i (14)$ & $0.0043$ & $0.0042$ & $0.0033$ & $0.0049$ &
 $0.0049$ & $0.0033$ & $0.0039$ & $0.0043$ \\
 $H^\pm W (8)$ & $0.18$ & $0.18$ & $0.19$ & $0.059$ &
 $0.20$ & $0.13$ & $0.041$ & $0.042$ \\
 $H^\pm W (14)$ & $1.24$ & $1.19$ & $1.15$ & $0.40$ &
 $1.02$ & $0.62$ & $0.26$ & $0.28$ \\
 $H^\pm t (8)$ & $0.31$ & $0.31$ & $0.31$ & $0.21$ &
 $0.26$ & $0.078$ & $0.078$ & $0.078$ \\
 $H^\pm t (14)$ & $1.97$ & $1.97$ & $1.97$ & $1.39$ &
 $1.69$ & $0.49$ & $0.49$ & $0.49$ \\
\hline
\end{tabular}
\end{center}
\caption{Cross sections (pb) for the charged Higgs production channels at the LHC with $\sqrt{s}=8-14$ TeV. \label{Table:CSHc}}
\end{table}

If we combine the two results, it is clear that the best candidate is $P'_5$, which shows the best features for both the production and the decay. Such strong production is triggered by the high value of the $H_2$ mass ($500$ GeV), which produces a charged Higgs $H^\pm$ ($400$ GeV) together with a vector boson $W^\pm$ ($\sim 80$ GeV) almost resonantly. This point has a very similar behaviour to the benchmarks $P_5$ and $P_6$ of Refs.~\cite{Basso:2012st,Basso:2013hs}, for which our proposed strategy can be easily applied. We can conclude that the portion of the parameter space in the neighborhood of $P'_5$ is a good candidate for a charged Higgs analysis in association with bosons.

\section{High masses and high $\tan\beta$}
\label{Sect:high-M-tanbeta}
The parameter space is severely constrained at high values of $M_{H^\pm}$ and $\tan\beta$. There is actually an interplay of three constraints that operate in this region. Below, we shall comment on these.

First of all, the electroweak precision data, in particular $T$ (or $\Delta\rho$), severely constrain the {\it splitting} of the second doublet. However, in the exact limit 
\begin{equation}
M_2=M_3=M_{H^\pm}\equiv M,
\end{equation}
the additional contributions to $T$ cancel. So this particular constraint can be evaded by such tuning of the heavy masses, which we will explore in the following.

There is a price to pay, the ``soft'' mass parameter must be carefully tuned, $\mu\simeq M$, due to the positivity and unitarity constraints.
In the limit when $M$ and $\mu$ both are large (compared to $M_1$) and $\tan\beta\gg1$, we find \cite{ElKaffas:2007rq}
\begin{subequations}
\begin{align}
\lambda_1&\simeq\frac{\tan^2\beta}{v^2}[c_1^2c_2^2M_1^2+(1-c_1^2c_2^2)M^2-\mu^2],  \label{Eq:lambda1}\\
\lambda_2&\simeq\frac{1}{v^2}[s_1^2c_2^2M_1^2+(1-s_1^2c_2^2)M^2], \label{Eq:lambda2} \\
\lambda_3&\simeq\frac{\tan\beta}{v^2}[c_1s_1c_2^2(M_1^2-M^2)]
+\frac{1}{v^2}[2M^2-\mu^2], \label{Eq:lambda3} \\
\lambda_4&\simeq\frac{1}{v^2}[s_2^2M_1^2+(c_2^2-2)M^2+\mu^2], \label{Eq:lambda4} \\
\Re\lambda_5&\simeq\frac{1}{v^2}[-s_2^2M_1^2+(\mu^2-c_2^2M^2)], \label{Eq:lambda5} \\
\Im\lambda_5&\simeq\frac{1}{v^2}c_2s_2(c_1+\tan\beta s_1)(M^2-M_1^2).
\end{align}
\end{subequations}

In the CP-conserving limit with $c_2=1$ and $H_3=A$, we have
\begin{subequations}
\begin{align}
\lambda_1&\simeq\frac{\tan^2\beta}{v^2}[c_1^2M_1^2+s_1^2M^2-\mu^2],  \label{Eq:lambda1bis}\\
\lambda_2&\simeq\frac{1}{v^2}[s_1^2M_1^2+c_1^2M^2], \label{Eq:lambda2bis} \\
\lambda_3&\simeq\frac{\tan\beta}{v^2}c_1s_1(M_1^2-M^2)
+\frac{1}{v^2}[2M^2-\mu^2], \label{Eq:lambda3bis} \\
\lambda_4&\simeq\frac{1}{v^2}(\mu^2-M^2), \label{Eq:lambda4bis} \\
\Re\lambda_5&\simeq\frac{1}{v^2}(\mu^2-M^2), \label{Eq:lambda5bis} \\
\Im\lambda_5&=0.
\end{align}
\end{subequations}

Tree-level unitarity roughly requires $|\lambda_i|<8\pi$ \cite{Ginzburg:2003fe}.
From Eqs.~(\ref{Eq:lambda4}) and (\ref{Eq:lambda5}) it follows that we must have $c_2\simeq1$ and $\mu\simeq M$.
Eqs.~(\ref{Eq:lambda1}) and (\ref{Eq:lambda2}) further require $|s_1|\simeq1$. The latter statement is made more precise by considering Eq.~(\ref{Eq:lambda3}), from which it follows that $c_1 \tan\beta M^2/v^2\leq{\cal O}(1)$, meaning that when $\tan\beta$ is large, $c_1$ must be very small, or $\alpha_1$ close to $\pm\pi/2$.

The positivity constraints can be written as \cite{Deshpande:1977rw}
\begin{subequations}
\begin{align} \label{Eq:positivity}
\lambda_1&>0, \quad \lambda_2>0, \\
\lambda_3&+\min[0,\lambda_4-|\lambda_5|]>-\sqrt{\lambda_1\lambda_2}.
\label{Eq:pos:lambda3}
\end{align}
\end{subequations}
The first of these equations, together with Eq.~(\ref{Eq:lambda1}), leads to $\mu\leq M$.
Then, $\min[0,\lambda_4-|\lambda_5|]=-2(M^2-\mu^2)/v^2$. By unitarity, the RHS of Eq.~(\ref{Eq:pos:lambda3}) can not be lower than $-8\pi$. Thus, we have a bound on $\tan\beta$, unless $c_1s_1$ vanishes:
\begin{equation}
c_1s_1 \tan\beta <\frac{M^2+8\pi v^2}{M^2-M_1^2},
\end{equation}
where we have approximated $\mu\simeq M$. On the other hand, if we set $s_1=1$ (but keep $\mu< M$), we find
\begin{equation}
\tan\beta<\sqrt{\frac{8\pi v^2}{M^2-\mu^2}}.
\end{equation}
Clearly, large values of $\tan\beta$ require either $c_1s_1\to0$ or $\mu\to M$.
For $\mu=0$, the cut-off is around $\tan\beta=5$--7, depending on the value of $M_{H^\pm}$ considered \cite{WahabElKaffas:2007xd}.
In addition, for a fixed value of $M_{H^\pm}$ there is a cut-off on $\tan\beta$ from $B\to\tau\nu$ and $B\to\tau\nu X$ decays \cite{Hou:1992sy}.

As stressed by Ref.~\cite{Eberhardt:2013uba}, there is an $M\gg v$ region, but our higher-dimensional parameter scan has some difficulty finding it. See, however, Fig.~\ref{Fig:alphas1-mh3-03}, where the case $M_2=M_{H^\pm}=600~\text{GeV}$ is excluded only by the ATLAS result on $R_{\gamma\gamma}$.

\section{Summary}

In this note, we have reviewed the status of the experimental and theoretical limits on a CP-violating version of the 2HDM type-II in view of the Moriond 2013 updates from the LHC. From the surviving parameter space, we have chosen some candidate points and we have checked the possibility of applying the selection strategy previously explained in \cite{Basso:2012st}. It turns out that, despite the considerable shrinking of the parameter space, there is still room for our proposed analysis. In practice, among our proposed benchmarks, a choice like $P'_5$ and similar configurations with $M_{H_2}\sim M_{H^\pm}+M_W$ give raise to a very good production cross section and decay rate. Hence, we persist with our suggestion for the next era of charged Higgs searches: after the discovery of the Higgs-like boson, the production and decay charged Higgs channels associated with bosons at hadron colliders deserve a special attention, because while the fermionic-associated channels are accompained by a huge background ($t\bar{t}$), the latter can be suppressed by a proper cutting strategy in the case of bosonic-associated channels.

\medskip
\noindent
{\bf Acknowledgement:}
We are grateful to the authors of Ref.~\cite{Eberhardt:2013uba} for clarifying discussions.



\begin{thebibliography}{99}
\bibitem{:2012gk}
  G.~Aad {\it et al.}  [ATLAS Collaboration],
  Phys.\ Lett.\ B {\bf 716} (2012) 1
  [arXiv:1207.7214 [hep-ex]].

\bibitem{:2012gu}
  S.~Chatrchyan {\it et al.}  [CMS Collaboration],
  Phys.\ Lett.\ B {\bf 716} (2012) 30
  [arXiv:1207.7235 [hep-ex]].

\bibitem{Djouadi:2005gi}
  A.~Djouadi,
  Phys.\ Rept.\  {\bf 457} (2008) 1
  [hep-ph/0503172].

\bibitem{Djouadi:2005gj}
  A.~Djouadi,
  Phys.\ Rept.\  {\bf 459} (2008) 1
  [hep-ph/0503173].

\bibitem{Gunion:1989we}
  J.~F.~Gunion, H.~E.~Haber, G.~L.~Kane and S.~Dawson,
  Front.\ Phys.\  {\bf 80} (2000) 1.

\bibitem{Basso:2012st}
  L.~Basso, A.~Lipniacka, F.~Mahmoudi, S.~Moretti, P.~Osland, G.~M.~Pruna and M.~Purmohammadi,
  JHEP {\bf 1211} (2012) 011
  [arXiv:1205.6569 [hep-ph]].

\bibitem{Lee:1973iz} 
  T.~D.~Lee,
  Phys.\ Rev.\ D {\bf 8}, 1226 (1973).
  
\bibitem{Shu:2013uua} 
  J.~Shu and Y.~Zhang,
  arXiv:1304.0773 [hep-ph].
  
\bibitem{Accomando:2006ga} 
  E.~Accomando, A.~G.~Akeroyd, E.~Akhmetzyanova, J.~Albert, A.~Alves, N.~Amapane, M.~Aoki and G.~Azuelos {\it et al.},
  hep-ph/0608079.
  
\bibitem{Khater:2003wq} 
  W.~Khater and P.~Osland,
  Nucl.\ Phys.\ B {\bf 661}, 209 (2003)
  [hep-ph/0302004].
  
\bibitem{ElKaffas:2007rq} 
  A.~W.~El Kaffas, P.~Osland and O.~M.~Ogreid,
  Nonlin.\ Phenom.\ Complex Syst.\  {\bf 10}, 347 (2007)
  [hep-ph/0702097 [HEP-PH]].

\bibitem{ATLAS:2013oma} 
  [ATLAS Collaboration],
  ATLAS-CONF-2013-012.
  
\bibitem{CMS-2013}
CMS talk at CERN, April 2013.

\bibitem{Guillelmo:2013cca} 
  G.~Gomez-Ceballos [CMS Collaboration],
  arXiv:1304.4194 [hep-ex].
  
\bibitem{ATLAS:2013nma} 
  [ATLAS Collaboration],
  ATLAS-CONF-2013-013.
  
\bibitem{ATLAS:2012ae} 
  G.~Aad {\it et al.}  [ATLAS Collaboration],
  Phys.\ Lett.\ B {\bf 710}, 49 (2012)
  [arXiv:1202.1408 [hep-ex]].
  
\bibitem{Chatrchyan:2012tx} 
  S.~Chatrchyan {\it et al.}  [CMS Collaboration],
  Phys.\ Lett.\ B {\bf 710}, 26 (2012)
  [arXiv:1202.1488 [hep-ex]].

\bibitem{Barroso:2013zxa} 
  A.~Barroso, P.~M.~Ferreira, R.~Santos, M.~Sher and J. P.~Silva,
  arXiv:1304.5225 [hep-ph].

\bibitem{Grinstein:2013npa} 
  B.~Grinstein and P.~Uttayarat,
  arXiv:1304.0028 [hep-ph].

\bibitem{Coleppa:2013dya} 
  B.~Coleppa, F.~Kling and S.~Su,
  arXiv:1305.0002 [hep-ph].

\bibitem{Chen:2013rba} 
  C.~-Y.~Chen, S.~Dawson and M.~Sher,
  arXiv:1305.1624 [hep-ph].
  
\bibitem{Eberhardt:2013uba} 
  O.~Eberhardt, U.~Nierste and M.~Wiebusch,
  arXiv:1305.1649 [hep-ph].
  
\bibitem{Hermann:2012fc} 
  T.~Hermann, M.~Misiak and M.~Steinhauser,
  JHEP {\bf 1211}, 036 (2012)
  [arXiv:1208.2788 [hep-ph]].

\bibitem{Basso:2013hs} 
  L.~Basso, A.~Lipniacka, F.~Mahmoudi, S.~Moretti, P.~Osland, G.~M.~Pruna and M.~Purmohammadi,
  arXiv:1301.4268 [hep-ph].

\bibitem{Ginzburg:2003fe} 
  I.~F.~Ginzburg and I.~P.~Ivanov,
  hep-ph/0312374.
  
\bibitem{Deshpande:1977rw} 
  N.~G.~Deshpande and E.~Ma,
  Phys.\ Rev.\ D {\bf 18}, 2574 (1978).
  
\bibitem{WahabElKaffas:2007xd} 
  A.~Wahab El Kaffas, P.~Osland and O.~M.~Ogreid,
  Phys.\ Rev.\ D {\bf 76}, 095001 (2007)
  [arXiv:0706.2997 [hep-ph]].
  
\bibitem{Hou:1992sy} 
  W.~-S.~Hou,
  Phys.\ Rev.\ D {\bf 48}, 2342 (1993).
  
\end{thebibliography}
\end{document}